\documentstyle[aps,epsf]{revtex}

\begin{document}

\author{T. A. S. Haddad{$^\ast$}, Angsula Ghosh and S. R. Salinas}
\address{Instituto de F\'{\i}sica, Universidade de S\~{a}o Paulo\\
Caixa Postal 66318, 05315-970, S\~{a}o Paulo, SP, Brazil}
\title{Tricritical behavior in deterministic aperiodic Ising systems}
\date{\today}
\maketitle

\begin{abstract}

We use a mixed-spin model, with aperiodic ferromagnetic exchange
interactions and crystalline fields, to investigate the effects of
deterministic geometric fluctuations on first-order transitions and
tricritical phenomena. The interactions and the crystal field parameters are
distributed according to some two-letter substitution rules. From a
Migdal-Kadanoff real-space renormalization-group calculation, which turns
out to be exact on a suitable hierarchical lattice, we show that the effects
of aperiodicity are qualitatively similar for tricritical and simple
critical behaviour. In particular, the fixed point associated with
tricritical behaviour becomes fully unstable beyond a certain threshold
dimension (which depends on the aperiodicity), and is replaced by a
two-cycle that controls a weakened and temperature-depressed tricritical
singularity.
\end{abstract}

The introduction of quenched disorder weakens (and sometimes eliminates)
first-order transitions and tricritical singularities in the phase diagram
of statistical models. In two dimensions, rigorous arguments show that any
amount of disorder completely eliminates first-order transitions in
ferromagnetic model systems \cite{aizen}. In three dimensions, approximate
real-space and perturbative renormalization-group analyses \cite{berker}, as
well as numerical simulations, indicate that a finite strength of disorder
is required to weaken first-order transitions and depress the tricritical
temperature. In particular, disordered versions of the two-dimensional
ferromagnetic $q$-state Potts model \cite{landau,cardy2} (for $q>4$, on the
square lattice, the uniform model displays a first-order transition) and the
Blume-Emery-Griffiths (BEG) model \cite{berkerfalicov,berker2,branco} (whose
uniform version displays tricritical and critical-end points), which have
been thoroughly investigated, are well adjusted to this scenario. It remains
unclear the important question of what are (if any) the universality classes
of the disorder-induced continuous transitions in these systems (see, for
example, a recent review by Cardy \cite{cardy}).

Instead of looking at the (presumably) more difficult problem posed by
fluctuations associated with quenched disorder, in the present publication
we consider the effects on first-order transitions of the geometric
fluctuations introduced by deterministic but aperiodically distributed
exchange interactions. For quenched disordered interactions, the Harris
criterion \cite{harris} indicates a change in the critical behaviour of
simple ferromagnets whenever the critical exponent associated with the
singularity of the specific heat of the underlying uniform model is
positive. According to a heuristic argument of Luck \cite{luck}, which has
been checked in a number of cases \cite[and references therein]{baake}, the
introduction of aperiodic interactions leads to an analogous criterion of
relevance of the geometric fluctuations on the critical (second-order)
behaviour. Some of us have recently shown that a similar criterion may be
exactly established for ferromagnetic Potts models on Migdal-Kadanoff
hierarchical lattices, with a layered distribution of exchange interactions
according to a class of two-letter substitution rules \cite{physa,salinas}.
For relevant geometric fluctuations, there appears a two-cycle of the
recursion relations in parameter space that gives rise to a new universality
class of (aperiodic) critical behaviour \cite{pre}. Along the lines of these
investigations, we introduce aperiodic interactions in a simple mixed-spin
model to analyze the effects on tricritical behaviour and first-order phase
boundaries. It should be mentioned that recent extensive Monte Carlo
calculations indicate that the phase transition of the $8$-state
square-lattice Potts model is indeed driven to second order by a layered
aperiodic distribution of exchange couplings \cite{berche}.

Besides the better known BEG model, another simple generalization of the
Ising model displaying first-order transitions and tricritical points is a
mixed-spin system, given by the Hamiltonian 
\begin{equation}
{\mathcal{H}}=-\sum_{(i,j)}J_{ij}\sigma_{i}S_{j}+\sum_{j}D_{j}S_{j}^{2},  
\label{hamilt1} 
\end{equation}
where $\sigma_{i}=\pm1$, for $i$ belonging to one sublattice of a bipartite
lattice, $S_{j}=\pm1$ or $0$, for $j$ belonging to the other sublattice, and
the first sum is over nearest-neighbor sites on different sublattices. The
description of this mixed-spin model demands a larger unit cell than the BEG
model in zero field. This extra difficulty, however, poses no problem to the
study of the model under a Migdal-Kadanoff real-space renormalization-group
approximation. Moreover, we need just two scaling fields, instead of three
as in the BEG model, to describe the even space of parameters. We then take
advantage of this model, and of the simplicity of the Migdal-Kadanoff
approximation (which turns out to be exact on a suitable hierarchical
lattice), to introduce aperiodic interactions for analyzing the effects of
geometric fluctuations on the main features of the phase diagram.

To obtain the Migdal-Kadanoff recursion relations, it is convenient to
rewrite Hamiltonian (\ref{hamilt1}) in the equivalent form 
\begin{equation}
{\mathcal{H}}=-\sum_{(i,j)}J_{ij}\sigma
_{i}S_{j}+\frac{1}{2}\sum_{(i,j)}D_{ij}\sigma
_{i}^{2}S_{j}^{2}.  \label{hamilt2}
\end{equation}
Many (mainly approximate) results are known for the uniform ($J_{ij}=J$,
$D_{ij}=D$) version of this model. On a honeycomb lattice, a star-triangle
transformation (summing over $S_{j}$) can be used to reduce the problem to
an exactly soluble spin-$1/2$ Ising model on a simple triangular lattice, in
which case, however, the temperature $(k_{B}T)$ versus ``anisotropy'' $(D/J)$
phase diagram presents only a line of continuous transitions \cite{lindberg}.
For the so-called union jack lattice, an exact solution can also be found for
a restricted range of parameters, by mapping the model onto an eight-vertex
problem \cite{horiguchi}. On a lattice of sufficiently high coordination, some effective-field \cite{kane1} and
self-consistent \cite{benayad} approximations, as well as a Bethe lattice
calculation \cite{natanael}, suggest the existence of a first-order boundary
that becomes a $\lambda $ line beyond a tricritical point. A detailed
Migdal-Kadanoff renormalization-group calculation \cite{quadros} predicts
the existence of a tricritical point on hypercubic lattices of dimension
$d\gtrsim 2.1$, which precludes the case of planar lattices. This is also
confirmed by renormalization-group calculations in momentum space\cite
{quadros}, at one-loop approximation, that do support the existence of the
tricritical point predicted by the Curie-Weiss version of the model. Monte
Carlo \cite{zhang} and numerical transfer matrix calculations \cite{buendia}
point out in this direction as well, with no indication of tricritical
phenomena in this mixed-spin model on a square lattice.

We now consider Hamiltonian (\ref{hamilt2}) and suppose that $J_{ij}$ (and
$D_{ij}$) may assume one out of two values, $J_{A}$ or $J_{B}$ ($D_{A}$ or
$D_{B}$), according to the sequence of letters generated by the iteration of
a substitution rule. In this paper, we work with two distinct binary rules, 
\begin{equation} \left( i\right) \quad A\rightarrow ABB,\quad B\rightarrow
AAA, \end{equation}
 and
\begin{equation}
\left( ii\right) \quad A\rightarrow AAB,\quad B\rightarrow AAA.
\end{equation}
For example, the iteration of the first rule leads to the following stages, 
\begin{equation}
A\rightarrow ABB\rightarrow ABBAAAAAA\rightarrow....
\end{equation}
Each rule is characterized by a substitution matrix, which relates the
number of letters $A$ and $B$ in one stage of the iteration with the
corresponding numbers in the previous stage. For the first rule, the
substitution matrix is given by 
\[
\mathbf{M}=\left( 
\begin{array}{ll}
1 & 3 \\ 
2 & 0
\end{array}
\right) , 
\]
with eigenvalues $\lambda_{1}=3$ and $\lambda_{2}=-2$. For the second rule,
we have 
\[
\mathbf{M}=\left( 
\begin{array}{ll}
2 & 3 \\ 
1 & 0
\end{array}
\right) , 
\]
with eigenvalues $\lambda_{1}=3$ and $\lambda_{2}=-1$. In a given stage of
the iteration, the fluctuation in the number of letters $A$ or $B$ relative
to the mean number behaves asymptotically as $N^{\omega}$, where $N$ is the
total number of letters in the sequence, and 
$\omega=\ln\left|
\lambda_{2}\right| /\ln\lambda_{1}$ is a wandering exponent.
We thus see
that the first rule, with an wandering
exponent $\omega=\ln2/\ln3$, gives rise to stronger geometric fluctuations
than the second one, with $\omega=0$, and should be more effective in
perturbing the critical behaviour. 
 
The Migdal-Kadanoff (MK) approximation on
a $d$-dimensional hypercubic
lattice turns out to be exact on the
hierarchical cell shown in figure 1. For
this type of cell, the MK scheme
corresponds to a decimation of the two
spins located along each bond, a
spin-$1/2$ and a spin-$1$, followed by the
moving and collapsing of $m$ such
bonds. There is also a relationship, $d=1+\ln m/\ln3$, between the number of
branches $m$ and the Euclidean
dimension $d$. Note that the renormalization
procedure amounts to a reverse
application of the substitution rule generated
by the aperiodic sequence.
Also, note that we can as well perform the more
usual trick of bond-moving
before decimation (which is also exact on a
suitable hierarchical lattice),
but the qualitative results should not depend
on this choice.
 
For the first aperiodic rule, $A\rightarrow ABB$,
$B\rightarrow AAA$, the MK
procedure yields the recursion relations 
\begin{eqnarray}
K_{A}^{\prime }=&&-\frac{m}{2}\ln \left\{ \exp (K_{A})+\exp (-K_{A})\cosh
\left( 2K_{B}\right) +\cosh \left( K_{B}\right)\exp 
\left( \frac{\Delta _{A}}{2}+\frac{\Delta
_{B}}{2}\right) \right\} \nonumber\\
&&+\frac{m}{2}\ln \left\{ \exp (K_{A})\cosh  \left(2K_{B}\right)
+\exp (-K_{A})+\cosh \left( K_{B}\right) \exp \left( \frac{\Delta
_{A}}{2}+\frac{\Delta _{B}}{2}\right) \right\},  \label{kal}
\end{eqnarray}

\begin{eqnarray}
K_{B}^{\prime }=&&-\frac{m}{2}\ln \left\{ \exp (K_{A})+\exp (-K_{A})\cosh
\left( 2K_{A}\right) +\cosh \left( K_{A}\right) \exp \left( \Delta
_{A}\right) \right\} \nonumber\\
&&+\frac{m}{2}\ln \left\{ \exp (K_{A})\cosh \left( 2K_{A}\right) +\exp(-K_{A})
+\cosh\left( K_{A}\right) \exp \left( \Delta _{A}\right) \right\}, \label{kbl}
\end{eqnarray}

\begin{eqnarray}
\Delta _{A}^{\prime }=&&-m\ln \left\{ \exp (-\Delta _{B}) \right. \nonumber\\
&&\times \left[ \exp (K_{A})+\exp (-K_{A})\cosh \left(2K_{B}\right) 
+\cosh \left( K_{B}\right) \exp \left( \frac{\Delta
_{A}}{2}+\frac{\Delta _{B}}{2}\right) \right]   \nonumber\\
&&\times \left[ \exp (K_{A})\cosh \left( 2K_{B}\right) +\exp (-K_{A})+\cosh
\left( K_{B}\right) \exp \left( \frac{\Delta _{A}}{2}+\frac{\Delta _{B}}{2}
\right) \right]   \nonumber\\
&&\times \left. \left[ 2\cosh (K_{A})\cosh (K_{B})+\exp \left( \frac{\Delta
_{A}}{2}+\frac{\Delta _{B}}{2}\right) \right] ^{-2}\right\}, \label{dal}
\end{eqnarray}

\begin{eqnarray}
\Delta _{B}^{\prime }=&&-m\ln \left\{ \exp (-\Delta _{A})\right. \nonumber\\
&&\times \left[ \exp (K_{A})+\exp (-K_{A})\cosh \left(
2K_{A}\right) +\cosh \left( K_{A}\right) \exp \left( \Delta _{A}\right)
\right] \nonumber\\
&&\times \left[ \exp (K_{A})\cosh \left( 2K_{A}\right)
+\exp (-K_{A})+\cosh \left( K_{A}\right) \exp \left( \Delta _{A}\right)
\right] \nonumber\\
&&\times \left. \left[ 2\cosh ^{2}(K_{A})+\exp \left(\Delta _{A}\right)
\right] ^{-2}\right\} ,  \label{dbl}
\end{eqnarray}
where $K_{A,B}=\beta J_{A,B}$ and $\Delta _{A,B}=\beta D_{A,B}$, with $\beta
=1/k_{B}T$. For all values of $m\geq 1$, these recursion relations have a
set of trivial fixed points, given by 
\begin{equation}
P_{+}\equiv (K_{A}^{*},K_{B}^{*},\Delta _{A}^{*},\Delta
_{B}^{*})=(0,0,-\infty ,-\infty ),  \label{p+}
\end{equation}
which is a sink of high-density paramagnetic phase (where the density is
related to the mean value $\left\langle S_{i}^{2}\right\rangle $, so that
low density means the predominance of spin $0$), 
\begin{equation}
P_{-}\equiv (K_{A}^{*},K_{B}^{*},\Delta _{A}^{*},\Delta
_{B}^{*})=(0,0,\infty ,\infty ),  \label{p-}
\end{equation}
which is a sink of low-density paramagnetic phase, 
\begin{equation}
O\equiv (K_{A}^{*},K_{B}^{*},\Delta _{A}^{*},\Delta _{B}^{*})=(0,0,0,0),
\label{b}
\end{equation}
corresponding to the high-temperature boundary between the paramagnetic
phases, 
\begin{equation}
F_{+}\equiv (K_{A}^{*},K_{B}^{*},\Delta _{A}^{*},\Delta _{B}^{*})=(\infty
,\infty ,-\infty ,-\infty ),  \label{f-}
\end{equation}
associated with a zero-temperature high-density ferromagnetic phase, and 
\begin{equation}
F_{-}\equiv (K_{A}^{*},K_{B}^{*},\Delta _{A}^{*},\Delta _{B}^{*})=(\infty
,\infty ,\infty ,\infty ),  \label{f+}
\end{equation}
with $\Delta _{A,B}^{*}/K_{A,B}^{*}=2$, which corresponds to a
zero-temperature low-density ferromagnetic phase.

For $m>1$ (which corresponds to $d>1$), there is a non-trivial fixed point
at $\Delta _{A}^{*}=\Delta _{B}^{*}=-\infty $, for $K_{A}^{*}=K_{B}^{*}$
finite. This fixed point, which we shall call $I$,  is associated with the
critical behaviour of the simple spin-$1/2$ Ising model, since fixing the
crystal field (biquadratic exchange) at $-\infty $ completely prevents the 
$S$ spins to assume $0$ values. In this spin-$1/2$ space, it should be 
pointed out that the recursion relations also present a \textit{two-cycle} 
(that is, a set of two fixed points of the second-iterate). As discussed in
a recent publication \cite{pre}, this two-cycle is associated with the new
universality class of the aperiodic ferromagnetic Ising model. Indeed, the
critical behaviour of the Ising model on the hierarchical lattice underlying
the MK approximation, and with aperiodic interactions according to the rule
$A\rightarrow ABB$, $B\rightarrow AAA$, is controlled by this two-cycle, the
singularity being weaker as compared with the uniform (periodic) model.

For $m\gtrsim3.33$ (numerical evidence points in fact to $10/3$),
corresponding to $d\gtrsim2.1$ (or $\ln10/\ln3=2.0959...$, as suggested by
the numerical calculations), there appear two novel non-trivial fixed
points. In the uniform case ($J_{A}=J_{B},$ $D_{A}=D_{B}$), Quadros and
Salinas \cite{quadros} have shown that one of them is a discontinuity fixed
point, associated with a first-order phase transition (according to an
application of the Nienhuis and Nauenberg criterion for the identification
of discontinuity fixed points). On the basis of a detailed analysis of the
connectivity of the flow lines in parameter space, the other fixed point of
the uniform model was shown \cite{quadros} to be associated with the
tricritical behaviour. In the uniform case, the discontinuity fixed point
displays a saddle character, with an attractive manifold emerging from the
zero-temperature high-density ferromagnetic trivial fixed point, and from
the tricritical fixed point (see figure 2, where a schematic view of the flows
is presented). The repulsive directions flow towards the sink of the
low-density paramagnetic phase, and towards the zero-temperature low-density
ferromagnetic trivial fixed point. The stable manifold is thus associated
with a first order line. The fully unstable tricritical fixed point is
connected with the Ising, $D=-\infty$, fixed point, giving rise to a
second-order boundary.

In the present aperiodic case, for $m\gtrsim3.33$, these two fixed points
still appear in the $K_{A}=K_{B}$, $\Delta_{A}=\Delta_{B}$, subspace, but
since the parameter space is four-dimensional we have to be careful to
generalize the overall picture of the last paragraph. In fact, we should pay
attention to the fact that now there are four scaling fields: the reduced
temperature, \textit{two} crystal fields (instead of only one, as in the
uniform case), and the \textit{strength} of the aperiodicity, as measured,
for example, by the ratio $r=J_{A}/J_{B}$. This last scaling field, which is
of completely different nature, is the determinant factor for the analysis
of stability of the fixed points against aperiodicity. If it turns out to be
relevant around a certain fixed point, it means that this fixed point cannot
be reached unless $r$ assumes a well-defined value, usually unity, so that
any amount of aperiodicity changes the critical behaviour controlled by this
particular fixed point. On the other hand, irrelevancy of this scaling field
suggests that, whatever the strength of the geometric fluctuations, this
fixed point continues to control the critical behaviour.

Consider now $m=3.4$ (dimension $d\approx2.1$). The fixed points are at
$K_{A}^{\ast}=K_{B}^{\ast}=1.5248...$, $\Delta_{A}^{\ast}=\Delta_{B}^{\ast
}=2.2118...$ (discontinuity fixed point in the uniform model) and
$K_{A}^{\ast}=K_{B}^{\ast}=1.1447...$, $\Delta_{A}^{\ast}=\Delta_{B}^{\ast
}=1.3953...$ (tricritical fixed point of the uniform case). The
linearization of the recursion relations (\ref{kal})-(\ref{dbl}), in the
neighborhood of this fixed point lead to the eigenvalues 
\[
\Lambda_{1}=4.3687...,\Lambda_{2}=0.7705...,\Lambda_{3}=-3.8520...,\Lambda
_{4}=-0.5825.... 
\]
The first two eigenvalues are the same as in the uniform case. The modulus
of the last two eigenvalues indicates the preservation of the saddle
character of this fixed point. The linearization about the second fixed
point yields the eigenvalues 
\[
\Lambda_{1}=4.1671...,\Lambda_{2}=1.2360...,\Lambda_{3}=-3.7346...,\Lambda
_{4}=-0.9194.... 
\]
Again, $\Lambda_{1}$ and $\Lambda_{2}$ assume the same values of the uniform
case. The modulus of the third eigenvalue is larger than $1$, so far
preserving the unstable character of the uniform tricritical fixed point.
The last eigenvalue, however, indicates an attractive direction towards this
fixed point, in contrast to the uniform model, in which case the tricritical
fixed point is fully unstable. However, we have already remarked that a
given characteristic fixed point associated with the uniform model still
controls the critical behaviour of the aperiodic system only if the strength
of the aperiodicity is an irrelevant scaling field. This is precisely what
is happening in this case. The modulus of $\Lambda_{4}$ asserts that the
tricritical fixed point associated with the uniform model can be reached,
even in the presence of aperiodicity.

An analysis of the flow lines of the recursion relations in parameter space
fully supports the idea that these two fixed points continue to perform
exactly the same functions in the aperiodic as well as in the uniform case.
In other words, the first one is indeed a discontinuity fixed point, and the
other one is associated with the tricritical behaviour. In a phase diagram
consisting of temperature, crystal fields and the aperiodicity ratio $r$,
there exists then a line of tricritical points extending along the $r$
direction. Finally, we note that $\Lambda_{3}$ and $\Lambda_{4}$ are
negative, for both fixed points. This is a common situation for these
aperiodic systems \cite{pre}, which reflects a flipping approximation to (or
furthering from) the fixed points, and is related to the discrete nature of
the renormalization procedure and to the existence of two competing energy
scales, $A$ and $B$.

For larger dimensions (larger values of $m$), this whole picture is changed.
Take, for example, $m=9$, corresponding to three dimensions. The
discontinuity fixed point of the uniform model is located at
$K_{A}^{\ast}=K_{B}^{\ast }=4.8427...$, $\Delta_{A}^{\ast}=\Delta_{B}^{
\ast}=9.0255...$. The tricritical fixed point is located at
$K_{A}^{\ast}=K_{B}^{\ast}=0.4514...$, $\Delta
_{A}^{\ast}=\Delta_{B}^{\ast}=0.2190...$. The linearization of the recursion
relations in the neighborhood of these fixed points lead to the eigenvalues 
\[
\Lambda_{1}=12.0154...,\Lambda_{2}=0.0032...,\Lambda_{3}=-10.5075...,\Lambda
_{4}=-0.0002..., 
\]
around the discontinuity fixed point (the first two eingenvalues are the
same as in the uniform case), and 
\[
\Lambda_{1}=9.1600...,\Lambda_{2}=2.5865...,\Lambda_{3}=-9.0707...,\Lambda
_{4}=-1.7413..., 
\]
around the second fixed point (again, the first two eingenvalues are the
same as in the uniform case). Note that this time the modulus of all of the
eigenvalues of the uniform tricritical fixed point are larger than unity,
which means that it is fully unstable against any aperiodic perturbations.
Hence, it cannot be reached whatever the strength of the geometric
fluctuations, except in the trivial, uniform case, $J_{A}=J_{B}$,
$D_{A}=D_{B}$. As in the case of the spin-$1/2$ critical fixed point (
$D_{A}=D_{B}=-\infty$), there also appears a two-cycle of the recursion
relations. This two-cycle is located at 
\[
(K_{A}^{\ast},K_{B}^{\ast},\Delta_{A}^{\ast},\Delta_{B}^{
\ast})_{1}=(0.3759...,4.5304...,0.1472...,8.0027...), 
\]
and 
\[
(K_{A}^{\ast},K_{B}^{\ast},\Delta_{A}^{\ast},\Delta_{B}^{
\ast})_{2}=(2.0653...,0.2772...,3.4221...,0.0905...). 
\]

As discussed in a previous publication \cite{pre}, we should now study the
behaviour of the \textit{second iterate} of the recursion relations around
any one of the points belonging to the two-cycle. From the linearization of
the second iterates of the recursion relations about these points, we have
the eigenvalues 
\[
\Lambda_{1}=114.3038...,\Lambda_{2}=82.0353...,\Lambda_{3}=5.1867...,\Lambda
_{4}=0.0014... . 
\]
Now there is at least one eigenvalue with modulus less than $1$, which
guarantees that the two-cycle is physically accessible. From a numerical
analysis of the connectivity of the flow lines in parameter space, we can
check that this two-cycle is indeed associated with the tricritical behaviour
(in analogy with the second-order transitions associated with the two-cycle
in the spin-$1/2$ subspace).

For $m=27$, corresponding to $d=4$, we have the same general features. The
uniform tricritical fixed point is fully unstable, and there appears a
two-cycle, which is presumably associated with a novel tricritical behaviour.
Numerical calculations point out that the two-cycle appears as a kind of
bifurcation of the uniform tricritical fixed point, at $m\approx3.45$
(corresponding to $d\approx2.13$). Further numerical work shows that, for a
given ratio $J_{A}/J_{B}$, the temperature that locates the system inside
the basin of attraction of the two-cycle (that is, the tricritical
temperature) is systematically lower as compared with the tricritical
temperature of the uniform model (and thus leads to a smaller first-order
region in the phase diagram). These features are also generally present in
disordered tricritical systems in three dimensions \cite{berkerfalicov}.

For the aperiodic rule $A\rightarrow AAB$, $B\rightarrow AAA$, with a
smaller wandering exponent ($\omega=0$ in comparison with $\omega=\ln2/\ln3$
for the previous sequence), we have a different set of recursion relations,
but the results are similar. In this case, the calculations show that the
uniform tricritical fixed point remains accessible (there is one eigenvalue
with modulus less than $1$) whatever the value of $m$. The smaller
eigenvalue around the tricritical fixed point approaches unity as $m$ goes
to infinity. Weaker geometric fluctuations are therefore unable to
relevantly perturb the system, whose multicritical behaviour remains
unchanged with respect to the uniform model.

In conclusion, we have studied the effects of deterministic aperiodicity on
a simple system that displays first and second-order transition lines and a
tricritical point. Using a simple Migdal-Kadanoff approximation, we show
that a certain class of non-random geometric fluctuations may change the
tricritical behaviour (by turning into a fully unstable node the fixed point
associated with the uniform tricritical behaviour). There is then a two-cycle
of the recursion relations that is shown to control the tricritical
behaviour. Numerical calculations indicate a depression of the tricritical
temperatures (as it used to happen in the presence of quenched disorder). In
spite of the limitations of the MK approximation (or, alternatively, the
artificiality of the hierarchical lattices in which it turns out to be
exact), the results of this investigation provide suggestions for the
analysis of multicritical behaviour in similar systems on realistic Bravais
lattices.

The authors wish to thank A. P. Vieira for useful discussions. This work has
been supported by the Brazilian agency FAPESP.

\newpage

\begin{figure}
\begin{center}
\epsfbox{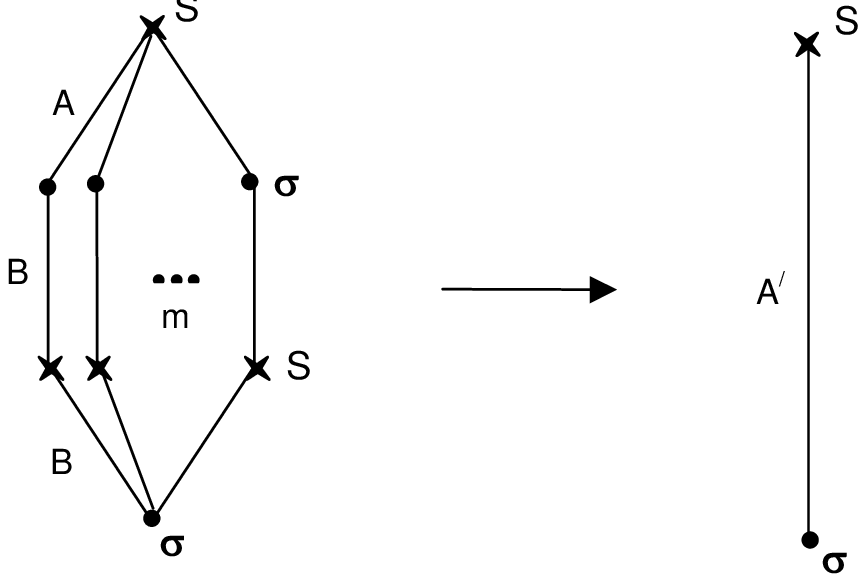}
\caption{Example of the hierarchical cell representing part of the
Migdal-Kadanoff decimation scheme. Crosses correspond to spins-1, and dots to
spins-1/2. By summing over the two internal spins of each of the $m$
branches, we obtain $K_{A}^{\prime}$ or $\Delta_{A}^{\prime}$.}
\end{center}
\end{figure}

\vspace{4.0truecm}

\begin{figure}
\begin{center}
\epsfbox{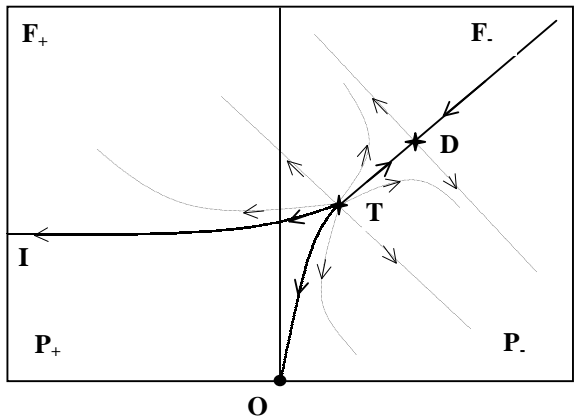}
\caption{Schematic view of the renormalization-group flows in parameter
space, for the uniform version of the mixed-spin Ising model, with $K$ in the
vertical axis, and $\Delta$ in the horizontal. $T$ is the tricritical fixed
point, and $D$ is the discontinuity one.}
\end{center}
\end{figure}

\end{document}